\documentclass[traditabstract]{aa} 
\usepackage{graphicx,natbib,longtable,txfonts,lscape}
\usepackage[switch]{lineno}
\newcommand{\ergs}{\>{\rm erg}\,{\rm s}^{-1}}
\newcommand{\kms}{$\rm{\,km \,s}^{-1}$}

\newcommand{\FR}{FRI{\sl{CAT}}}
\newcommand{\FRo}{FR0{\sl{CAT}}}
\newcommand{\sFR}{sFRI{\sl{CAT}}}
\usepackage{color}

\begin{document}

   \title{FR0{\sl{CAT}}: a FIRST catalog of FR~0 radio galaxies.}

   \subtitle{}

   \author{R.~D. Baldi\inst{1}
\and
A. Capetti\inst{2}
          \and
          F. Massaro\inst{3}
          }

   \institute{Department of
     Physics and Astronomy, University of Southampton, Highfield, SO17 1BJ, UK
\and INAF-Osservatorio Astrofisico di Torino, via Osservatorio 20,
     10025 Pino Torinese, Italy, \and
Dipartimento di Fisica, Universit\`a degli
     Studi di Torino, via Pietro Giuria 1, 10125 Torino, Italy}
   \date{}

  \abstract {With the aim of exploring the properties of the class of FR~0
    radio galaxies, we selected a sample of 108 compact radio sources, called
    FR0{\sl{CAT}}, by combining observations from the NVSS, FIRST, and SDSS
    surveys. We included in the catalog sources with redshift $\leq 0.05$,
    with a radio size $\lesssim$ 5 kpc, and with an optical spectrum
    characteristic of low-excitation galaxies. Their radio luminosities at 1.4
    GHz are in the range $10^{38} \lesssim \nu L_{1.4} \lesssim 10^{40}
    \ergs$. The FR0{\sl{CAT}} hosts are mostly (86\%) luminous ($-21 \gtrsim
    M_r \gtrsim -23$) red early-type galaxies with black hole masses $10^8
    \lesssim M_{\rm BH} \lesssim 10^9 M_\odot$. These properties are similar
    to those seen for the hosts of FR~I radio galaxies, but they are on average
    a factor $\sim$1.6 less massive.

    The number density of FR0{\sl{CAT}} sources is $\sim$5 times
    higher than that of FR~Is, and thus they represent the dominant
    population of radio sources in the local Universe. Different
    scenarios are considered to account for the smaller sizes and
    larger abundance of FR~0s with respect to FR~Is. An age-size scenario that
    considers FR~0s as young radio galaxies that will all eventually
    evolve into extended radio sources cannot be reconciled with the
    large space density of FR~0s. However, the radio activity
    recurrence, with the duration of the active phase covering a wide
    range of values and with short active periods strongly favored
    with respect to longer ones, might account for their large
    density number. Alternatively, the jet properties of FR~0s might
    be intrinsically different from those of the FR~Is, the former
    class having lower bulk Lorentz factors, possibly due to lower
    black hole spins.

Our study indicates that FR~0s and FR~I/IIs can be interpreted as two
extremes of a continuous population of radio sources that is characterized by a broad
distribution of sizes and luminosities of their extended radio emission, but
shares a single class of host galaxies.}  

\maketitle

\section{Introduction}

The widespread presence of compact radio sources at the center of
early-type galaxies (ETG) has been recognized in the 1970s
\citep{ekers73}. Later studies (\citealt{wrobel91b}, but see also
\citealt{sadler84} and \citealt{slee94}) performed deeper radio
surveys and found that $\sim$ 30\% of the nearby ETGs are detected
above $\sim$ 1 mJy level, reaching luminosities as low as
$\sim 2 \times 10^{19}$ W Hz$^{-1}$ (i.e., $\sim 3 \times 10^{35}$ erg
s$^{-1}$). The vast majority of these sources are unresolved at
3-5$\arcsec$ resolution, indicating that they are confined within a
region smaller than some kpc, and they often show short jet-like
features and no large-scale jets \citep{nagar05}. The origin of the
radio emission, whether due to star formation or to an active galactic
nucleus (AGN), has not always been identified unambiguously,
especially at very low radio fluxes ($<$ mJy level,
\citealt{bonzini13}). Furthermore, even for objects with a confirmed
AGN origin, it is unclear whether these low-power radio galaxies (RGs)
in nearby ETGs are truly the scaled-down versions of more powerful
radio AGN \citep{ho99} and how the lack of very extended radio
emission in sources with weak jets can be explained
\citep{fabbiano89}.  A detailed analysis of the link between the radio
source and its host requires additional observational data, mainly
optical imaging and spectroscopy.

Compact radio sources have been little studied by the radio community,
which reasonably prefers to explore the radio morphology and
properties of brighter RGs, such as the Third Cambridge Catalog of
Radio Sources (3C, \citealt{edge59,bennett62}) and its successors. The
high-flux limit sets for these samples does not favor the inclusion of
weak compact RGs. Fortunately, the advent of large-area
multiwavelength surveys opens up the opportunity to set the studies of
compact radio sources on strong statistical foundations. In
particular, these surveys allow us to identify large numbers of radio
sources, to obtain spectroscopic redshifts, and to determine the
properties of their hosts. Several studies (e.g.,
\citealt{best05b,baldi10b,best12,mingo16,miraghaei17}) have used the
extensive available multi-frequency information to analyze the
properties of the population of low-redshift low-power radio AGN. In
particular, \citet{best12} defined diagnostics to isolate galaxies in
which the radio emission is produced by an active nucleus.

\citet{baldi09} studied the radio properties of miniature radio
galaxies (named Core galaxies), which are known to have nuclei as the
scaled-down version of nuclei of the Fanaroff-Riley type~I galaxies,
FR\ Is, of the 3C sample \citep{balmaverde06core}. Nevertheless, the
former show a different radio behavior because they are smaller and
more core-dominated and consequently have a much lower extended radio
emission. Therefore, as a further analysis, we moved our focus to
large radio surveys to confirm the presence of a population of radio
sources that would show such a radio peculiarity, similarly to the
Core galaxies. The result emerging from these studies \citep{baldi09,
  baldi10b} is that the majority of radio AGN are unresolved (or
barely resolved) at the 5$\arcsec$ resolution of the FIRST
survey\footnote{Faint Images of the Radio Sky at Twenty centimeters
  survey \citep{becker95,helfand15}.} and radio weak, as expected
based on the local radio luminosity function (e.g.,
\citealt{best05b,pracy16}). This is in contrast with the results from samples
selected at higher flux limits where most radio sources are extended
and belong to the FR~I or FR~II classes \citep{fanaroff74}. The
general lack of extended radio structures suggests a definition of
these compact sources as FR~0 \citep{ghisellini11,baldi15}.

\citet{baldi15} presented the results of a pilot program of
high-resolution ($\sim 0\farcs2$) radio imaging of a small sample of
compact sources. They found that these can be associated with both
radio-quiet and radio-loud AGN; this second group, the genuine FR~0s,
are located in red massive ($\sim 10^{11} M_\odot$) ETGs with high BH
masses ($\gtrsim 10^8 M_\odot$) and are spectroscopically classified
as low-excitation galaxies (LEG). These are all characteristics
typical of FR~I RGs. They also lie on the correlation between radio
core power and [O~III] line luminosity defined by FR~Is, which is an
indication of a common mechanism for the production of radio and
ionizing radiation. This also rules out strong effects from Doppler
beaming and projection effects. However, they are unresolved at
sub-kpc resolution, or show jet-like radio structures on a scale of
1-3 kpc.  FR~0s are more core dominated (by a factor of $\sim$30) than
FR~Is \citep{baldi15}. In summary, the only substantial difference
between compact FR~0 and extended FR~I radio sources is the deficit of
extended radio emission.

Since the radio selection of compact radio galaxies carried out by
\citet{baldi10b} corresponds to an optical selection, that is, red
massive ETGs with an LEG optical spectrum, we adopted these radio and
spectrophotometric characteristics to define our FR~0 class of
RGs. However, a more heterogeneous population of compact radio sources
are present in the local Universe, which includes radio-quiet AGN, compact
steep-spectrum sources, and blazars (see \citealt{sadler14}), sources
that are not investigated in this work.

The radio sample selected by \citet{best05b} at 1.4 GHz mostly includes
low-power compact RG, which are morphologically consistent with an FR~0
definition. Similarly, at higher radio frequencies, the FR~0s appear
to be the dominant population \citep{sadler14,whittam16,whittam17}.

This work is the third of a series of papers aimed at studying the
low-power RGs in the local Universe: the first was about FR~Is
\citep{capetti17a} and the second about FR~IIs
\citep{capetti17b}. Here, we focus more on the properties of
the FR~0s by extricating a sample of compact radio sources from the
catalog of radio AGN defined by \citet{best12}. The properties of
  the FR~0s can then be compared with those of the FR~I RG
  selected from the same original sample reported by \citet{capetti17a},
  requiring an edge-darkened radio morphology and extending at least
  to a 30 kpc radius (the \FR\ sample) or between 10 and 30 kpc (the
  \sFR\ sample).

This paper is organized as follows. In Sect.\ 2 we present the
selection criteria of the FR~0 sample. The radio and optical
properties of the FR~0 catalog are presented in Sect.\ 3 and discussed
in Sect. 4. After a summary, our conclusions are drawn in Sect.\ 5.

Throughout the paper we adopt a cosmology with $H_0=67.8 \, \rm km \, s^{-1} \,
Mpc^{-1}$, $\Omega_{\rm M}=0.308$, and $\Omega_\Lambda=0.692$ \citep{ade16}.  
\section{Sample selection}
\label{sample}

\begin{figure*}
\includegraphics[width=9.2cm]{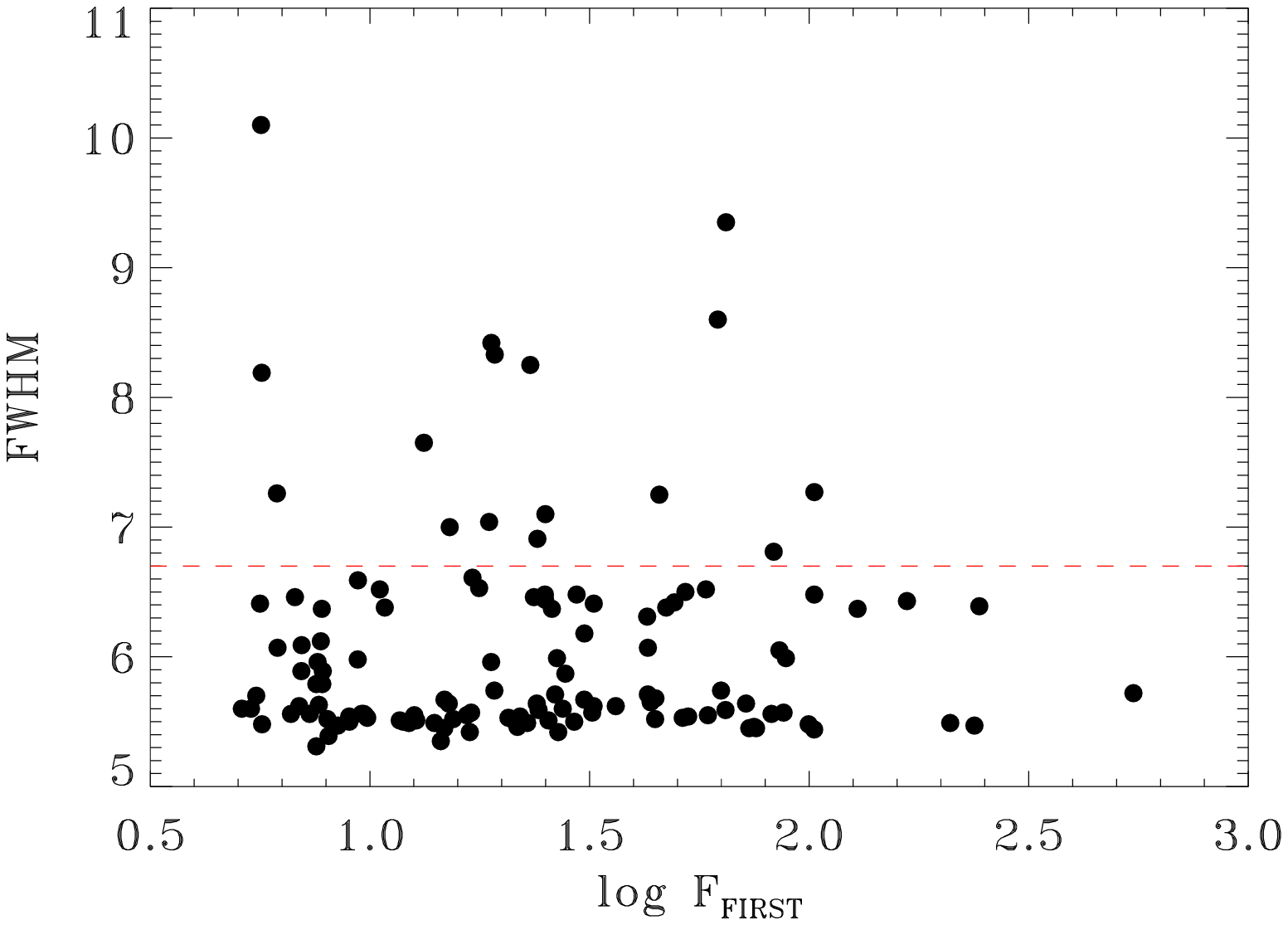}
\includegraphics[width=9.2cm]{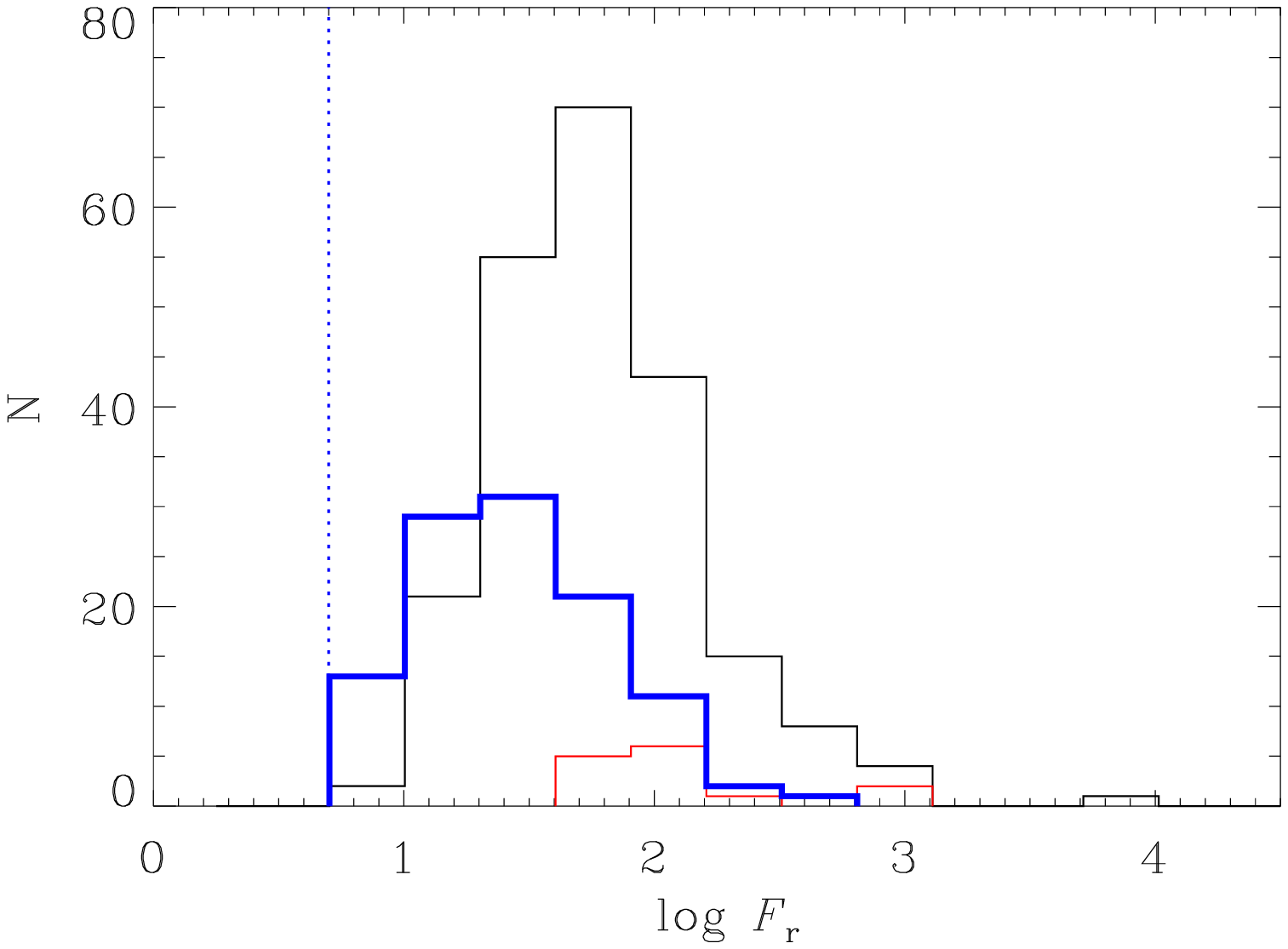}
\includegraphics[width=9.2cm]{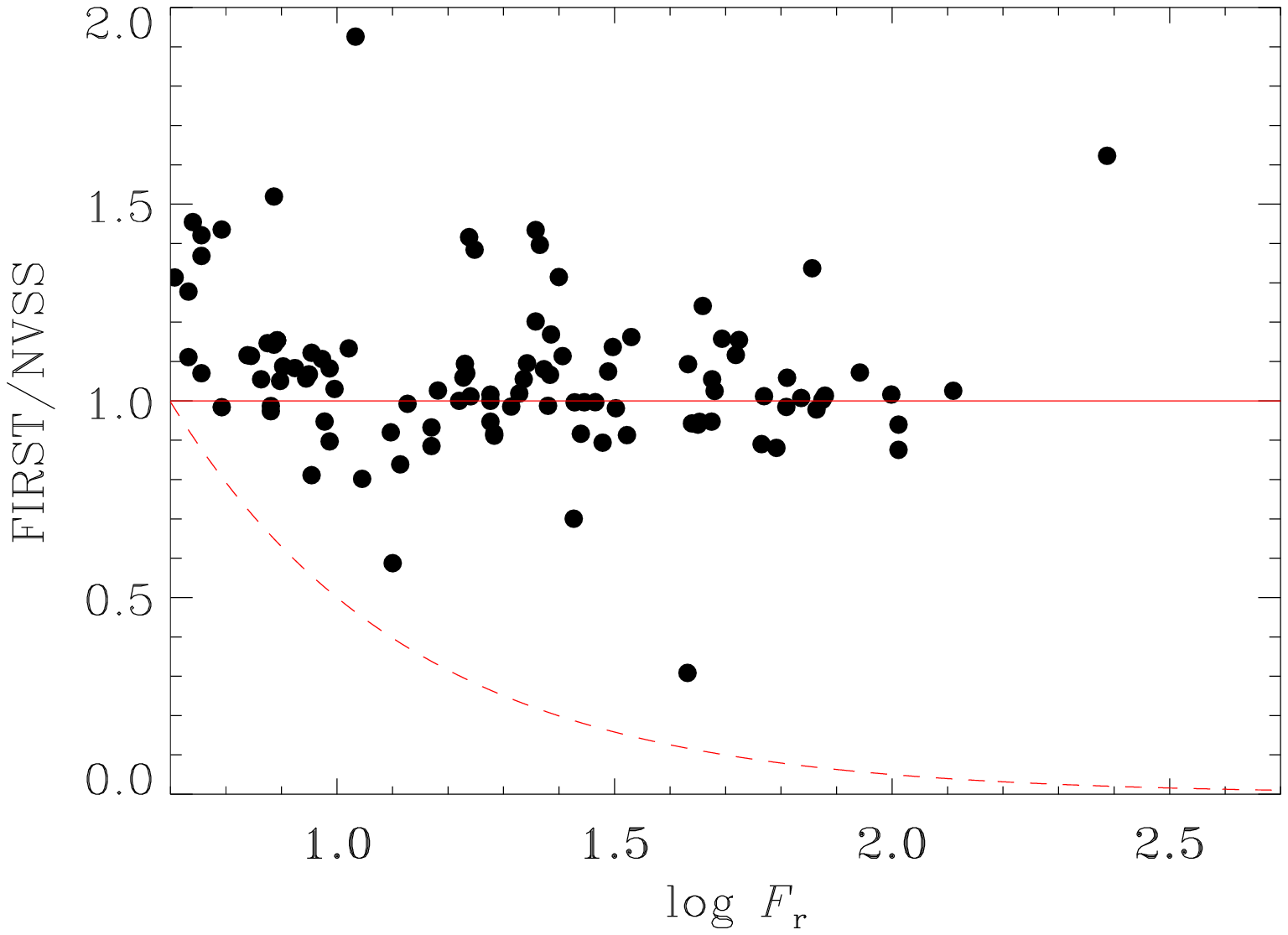}
\includegraphics[width=9.2cm]{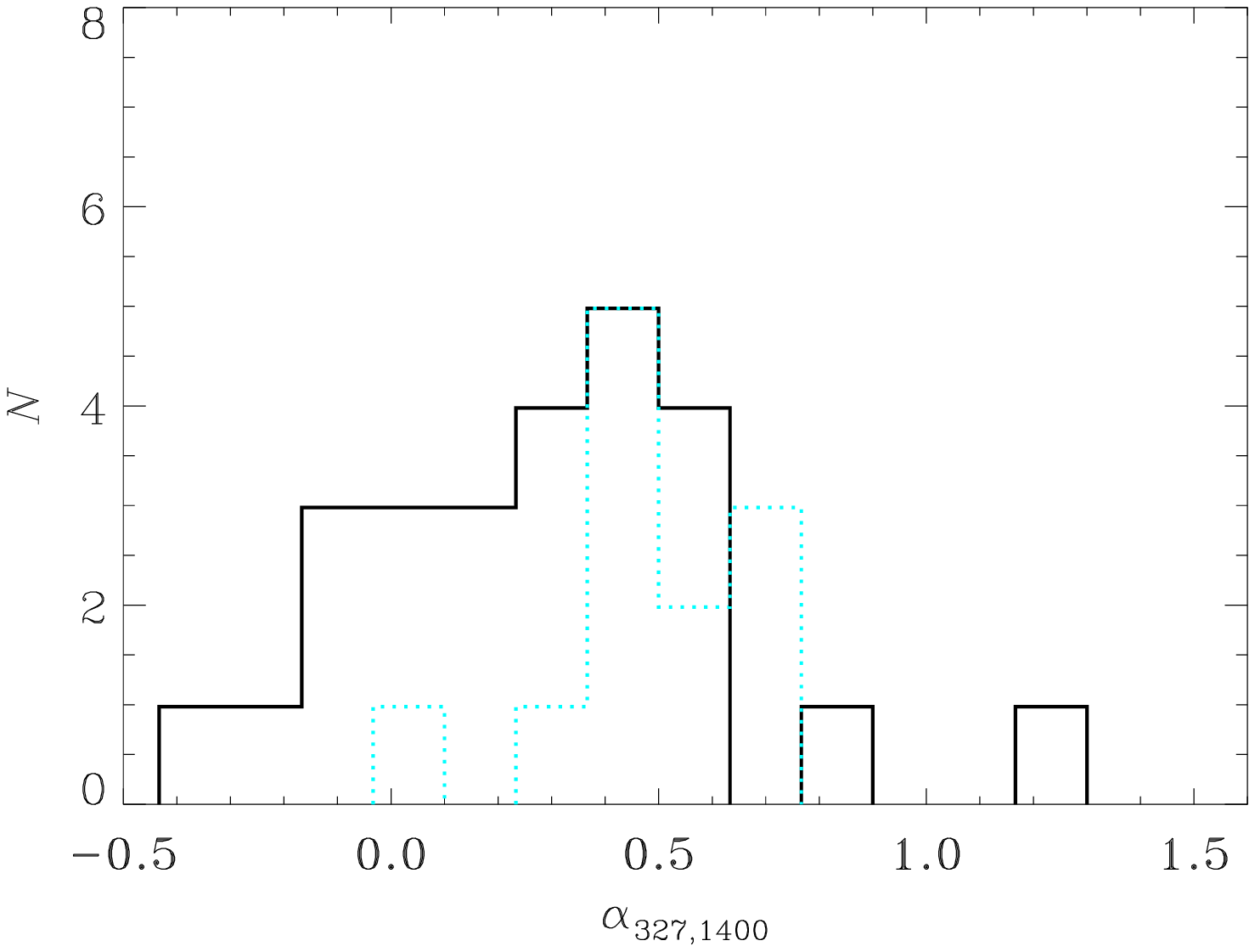}
\caption{Top left panel: measured major axis (FWHM) in arcsec versus the FIRST
  flux density for sources that do not show clearly resolved radio
  emission; the red dashed line marks the adopted limit of 6\farcs7
  for the inclusion in the \FRo\ sample. Top right panel: distribution of
  the NVSS fluxes (left) of the 108 \FRo\ (blue), the 219 \FR\
  (black), and the 14 \sFR\ sources (red). The vertical dotted line
  marks the 5 mJy limits of the BH12 sample. Bottom left panel: ratio
  between the FIRST and NVSS fluxes versus the NVSS flux density for
  the \FRo\ sources; the dashed curve indicates the upper
    boundary of the region in which objects are located that pass the NVSS flux
    threshold, but were not selected due to the 5 mJy minimum FIRST flux
    requirement because of a large contribution of
    resolved emission.  Bottom right panel: radio
  spectral index between 327 and 1400 MHz; the dashed cyan histogram
  represents the upper limits.}
\label{hist}
\end{figure*}

We seek for FR~0 RGs in the sample of 18,286 radio sources built
by \citet{best12} (hereafter the BH12 sample) by limiting the search
to a subsample of objects in which, according to these authors, the
radio emission is produced by an active nucleus based on different
diagnostics (i.e., SDSS spectra and radio vs. host photometry). They
cross-matched the optical spectroscopic catalogs produced by the group
from the Max Planck Institute for Astrophysics and The Johns Hopkins
University \citep{bri04,tre04} based on data from the data release 7
of the Sloan Digital Sky Survey (DR7/SDSS,
\citealt{abazajian09}),\footnote{Available at {\tt
    http://www.mpa-garching.mpg.de/SDSS/}.} with the National Radio
Astronomy Observatory Very Large Array Sky Survey (NVSS;
\citealt{condon98}) and FIRST, adopting a radio flux density limit of
5 mJy in the NVSS. The catalog includes mostly RLAGN, with a small
fraction ($\sim$10\%) of a possible radio-quiet AGN contribution
\citep{baldi10b}.

We initially selected the sources with 1) redshift $z \leq 0.05$ to optimize
the spatial resolution, 2) a maximum offset of $2\arcsec$ of the radio sources
from the optical center, and 3) a minimum FIRST flux of 5 mJy. The last
constraint is related to the possibility of an accurate size measurement (see
below). One hundred ninety-one sources passed this selection criteria.

We visually inspected the FIRST images of these sources and discarded
the objects with clearly extended radio emission. We gathered the
measured sizes of the remaining sources and preserved those in which the
deconvolved size is smaller than 4\arcsec, that is, sources in which the observed
major axis is smaller than 6\farcs7 (Fig. \ref{hist}, top left
panel). At $z=0.05$ this corresponds to $\sim$5 kpc, that is, to a radius
of 2.5 kpc.

\begin{figure*}
\includegraphics[width=9.2cm]{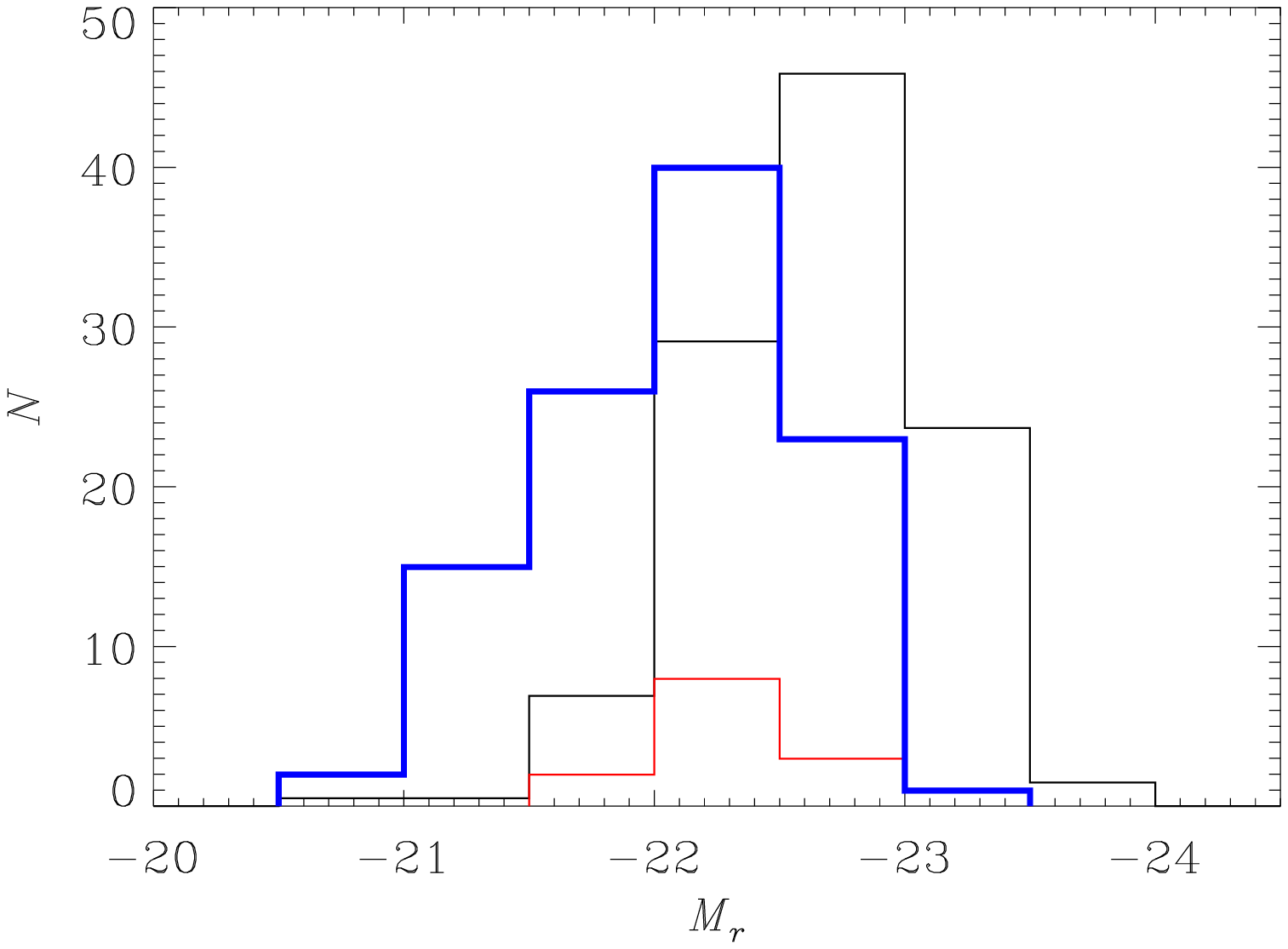}
\includegraphics[width=9.2cm]{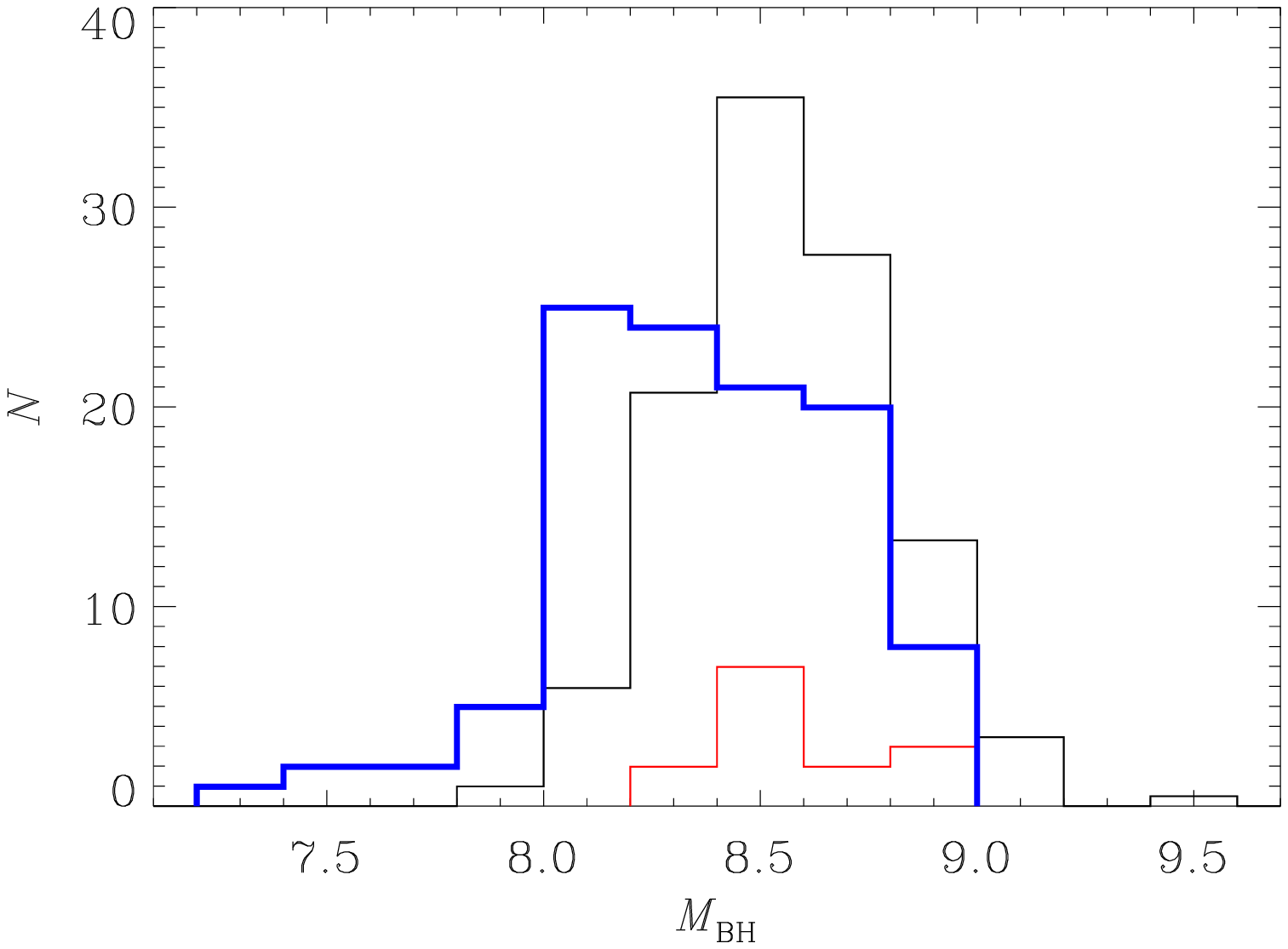}
\caption{Distributions of the $r$ -band absolute magnitude (left) and black
  hole masses (right). The \FR\ histograms are all scaled by the relative
  number of FR~I and FR~0, that is, by 108/219. Colors as in Fig. \ref{hist}.}
\label{mhist}
\end{figure*}

\begin{figure*}
\includegraphics[width=9.2cm]{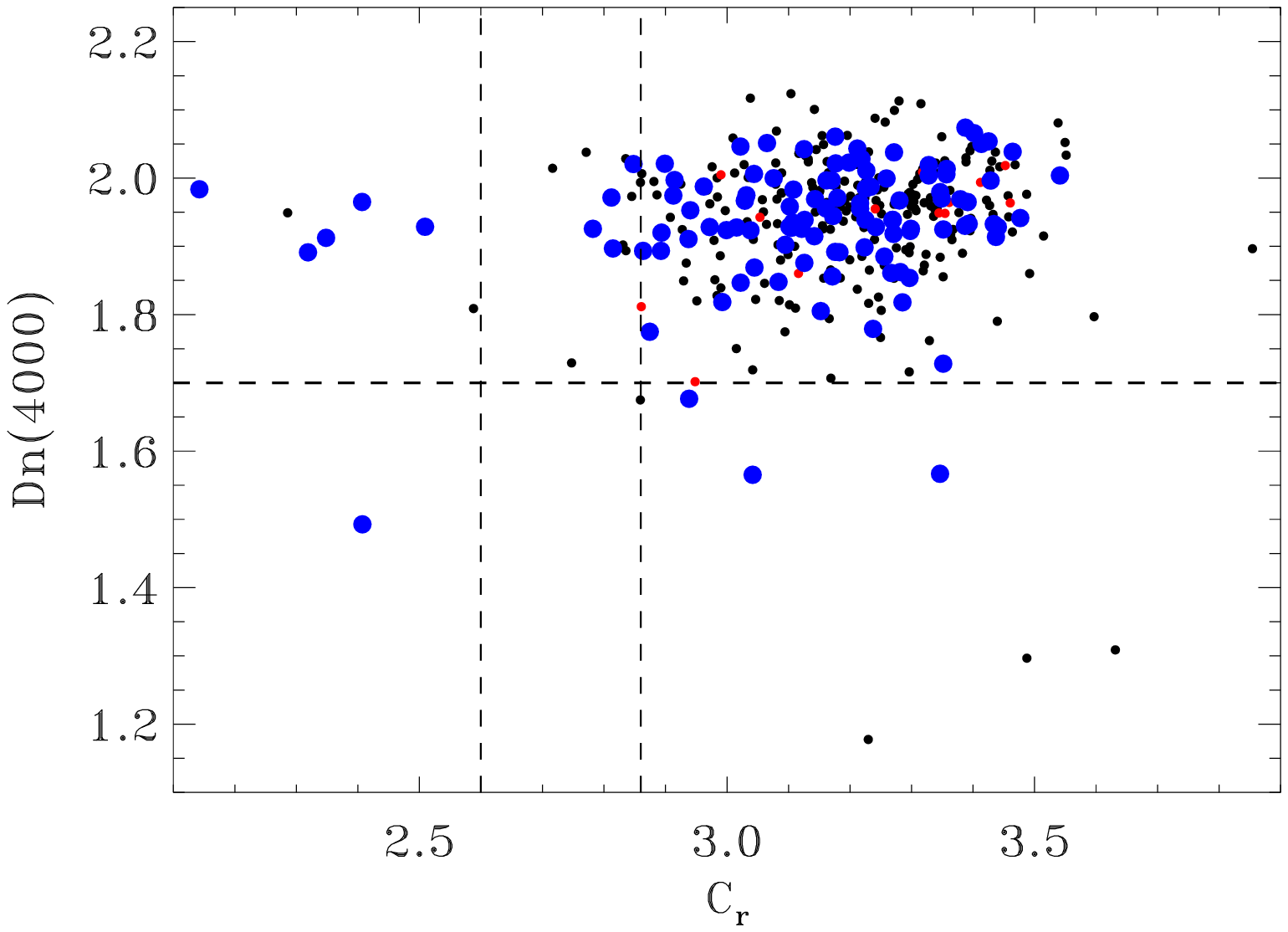}
\includegraphics[width=9.2cm]{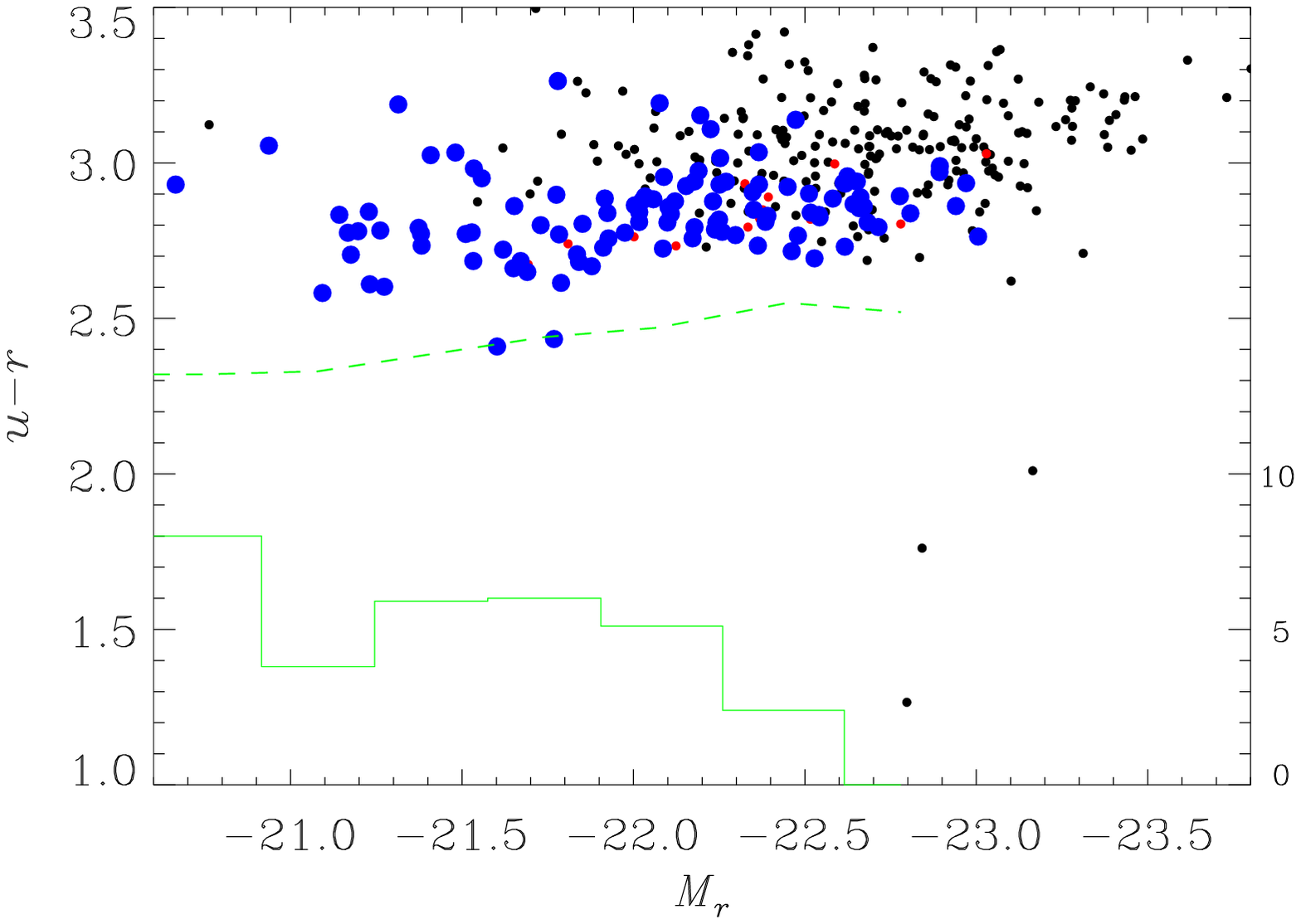}
\caption{Left panel: concentration index $C_r$ vs Dn(4000) index;
  right panel: absolute $r$-band magnitude, $M_r$, vs. $u-r$ color of the
  \FRo\ (blue), \FR\ (black), and \sFR\ (red) sources. The green histogram on
  the bottom shows the percentage of blue ETGs (scale on the right axis) from
  \citet{schawinski09}. The dashed line separates the blue ETGs from the
  red sequence, following their definition. Colors as in Fig. \ref{hist}.}
\label{crdn}
\end{figure*}

All selected galaxies are characterized by a spectrum typical of LEG,
with only four exceptions that have a high-excitation spectrum. Since
the main aim of this project is the comparison between FR~0s and
FR~Is, which typically have an LEG spectrum, we limit our analysis to
the LEG compact radio sources. The resulting sample, to which we refer
as \FRo, is formed by 108 FR~0s, whose main properties are presented
in Table \ref{tab}.

The size limit estimated above applies only to a source with a Gaussian shape.
This is unlikely to be the case and this question thus requires a more detailed
treatment. To estimate a reliable limit to the size of FR~0s, we
considered two possibilities: they are 1) compact doubles, or 2)
small FR~Is. For the first case we measured the FWHM of a simulated source
formed by two unresolved components of equal flux; the elliptical Gaussian
major axis exceeds the above threshold when their separation exceeds 3\farcs2.

The second case is more complex because of the variegated morphologies
shown by FR~Is. We explored it by simulating the images of small
edge-darkened radio sources by using the sample of 14 FR~Is (named
sFR~Is) selected in \citet{capetti17a} as a starting point. These
sources have a redshift $z<0.05$ and at least one jet that extends
between 10 and 30 kpc from the host. We simulated the appearance of
even smaller FR~Is by reducing the angular scale of their radio images
by a factor 3; this was obtained with a Gaussian smoothing and a
rebinning of the images. We then measured the FWHM of the resulting
images. Only 3 of these 14 sources would be considered as compact with
the angular size threshold adopted for the FR~0s selection. In the
original images the jets of these FR~Is extend by between 12 and 15
kpc, that is, 4-5 kpc with the reduced angular scale. The 11 remaining
sources, having a median size of $\sim$20 kpc, appear to be resolved.

These simulations suggest that a radio source with either a double or
an edge-darkened core-jet(s) morphology, located at $z<0.05$, will be
cataloged as FR~0 with our criteria only when its size does not exceed
$\sim$ 5 kpc.

Recently, \citet{miraghaei17} also selected compact RGs from the BH12
sample.  They selected single-component FIRST sources, brighter than
40 mJy, in the redshift range 0.03$>$z$>$0.1. Their criteria were
different from our selection, which leads to a marginal overlap
between the two samples, with only seven objects in common.

\section{Hosts and radio properties of FR~0s}
\label{hosts}

The NVSS radio flux, $F_r$, distribution of the \FRo\ (see
  Fig. \ref{hist}, top right panel) spans from the sample limit (5 mJy) to
  $\sim$ 400 mJy and peaks at $\sim$20-30 mJy. The presence of such a peak
  indicates that the selected sample is incomplete at a flux higher than the
  original selection threshold. This is due to the additional requirement of a
  minimum FIRST flux, $F_{\rm FIRST} >$ 5 mJy. Objects close to the 5 mJy NVSS
  flux limit might not reach the minimum $F_{\rm FIRST}$ value if there is a
  sufficient amount of extended emission resolved out in these higher
  resolution images. Nonetheless, the majority of the \FRo\ shows a ratio
between the FIRST and NVSS flux densities included in the range 0.5 - 1.5
(Fig. \ref{hist}, bottom left panel). This indicates that
  there is no significant amount of extended low-brightness radio emission
  associated with FR~0s in general.

Similarly to the \FR\ sample, the optical selection of the \FRo\ does
not introduce a significant incompleteness because most of the hosts
have $15.5 < r < 13$, which is the $r$-band magnitude range where the redshift
completeness of the SDSS is $\sim$90\% \citep{montero09}.

The radio spectral index can be measured for the 38 \FRo\ sources that
fall into the area covered by the 327 MHz Westerbork Northern Sky
Survey (WENSS, \citealt{rengelink97}). The 26 \FRo\ detected by the
WENSS show a broad distribution of $\alpha_{327,14000}$
(Fig. \ref{hist}, bottom right panel); 20 of them are flat spectrum
sources, $\alpha_{327,14000} < 0.5$, and to this category, we can add
seven objects undetected at the 18 mJy flux limit of the WENSS. This
indicates that the radio emission in \FRo\ sources is generally
dominated by a flat spectrum core, similar to the FR~0s studied by
\citet{baldi15}.

The absolute magnitude distribution of the FR~0s hosts covers the range
$-21 \lesssim M_r \lesssim -23$ (see Fig. \ref{mhist}, left panel). Their
black hole (BH) masses (Fig. \ref{mhist}, right panel), estimated from the
stellar velocity dispersion taken from the MPA-JHU spectral analysis of
  the DR7 SDSS data and the relation of \citet{tremaine02}, are in the range
$8.0 \lesssim \log M_{\rm BH} \lesssim 9.0 M_\odot$, with only ten objects
having $M_{\rm BH} \leq 10^8 M_\odot$. In the two panels we report the analogous
distributions for the \FR\ sources; FR~Is are associated with slightly
brighter sources and more massive BHs. The differences in the medians are
$\Delta M_r = 0.59$ and $\Delta \log M_{\rm BH} =0.20$, respectively,
corresponding to a factor $\sim$1.6 in both cases. The comparison with
\sFR\ leads to similar results, with $\Delta M_r = 0.35$ and $\Delta M_{\rm
  BH} =0.16$, that is, a factor $\sim$1.4. All these differences have a
statistical significance of at least 95\% according to the Kolmogoroff-Smirnov
test.

\begin{figure}
\includegraphics[width=9.2cm]{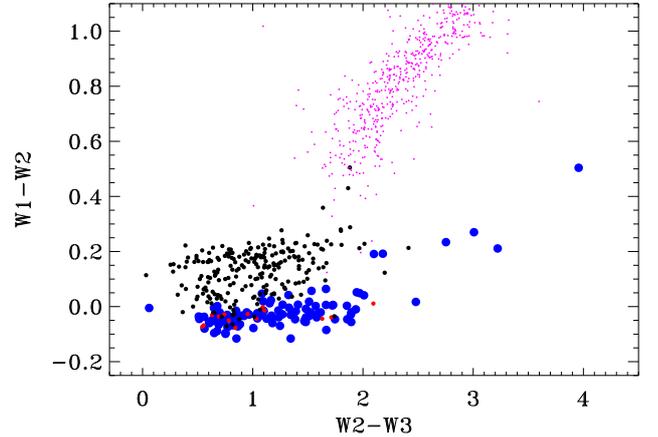}
\caption{{\em{WISE}} mid-IR colors of the \FRo\ hosts compared to the colors of
  \FR\ (black) and \sFR (red). We also show the region occupied by the
       {\em{Fermi}} blazars (purple dots).}
\label{wise}
\end{figure}

\begin{figure}
\includegraphics[width=9.0cm]{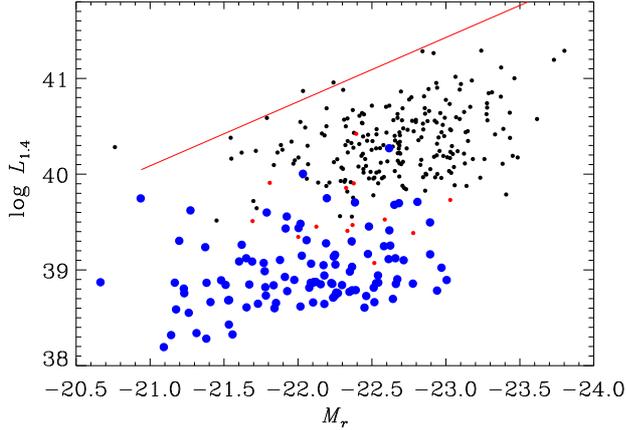}
\caption{Radio luminosity (NVSS) vs. host absolute magnitude , $M_r$, for
  \FRo\ (blue), \FR\ (black), and \sFR\ (red) sources. The solid line shows
  the separation between FR~I and FR~II reported by \citet{ledlow96}, to which
  we applied a correction of 0.34 mag to account for the different magnitude
  definition and the color transformation between the SDSS and Cousin
  systems.}
\label{ledlow}
\end{figure}

Similarly to what is seen in the \FR, the vast majority of the \FRo\
hosts are red ETGs, based on the values of the concentration $C_r$
\citep{strateva01} and spectroscopic Dn(4000) \citep{balogh99}
indices, see Fig. \ref{crdn}, left panel. Their redness is confirmed
by the photometric $u-r$ color, measured over the whole galaxy (see
Fig. \ref{crdn}, right panel), with only two galaxies located close to
the boundary between blue and red ETGs.

A small fraction of FR~0s departs from the general behavior.  These
are galaxies with $M_{\rm BH} \leq 10^8 M_\odot$, $C_r < 2.6$, or
Dn(4000) $< 1.7$. While this is due to spurious effects for some (in
two cases the low $C_r$ is due to a nearby star or a companion
galaxy), 15 galaxies appear to be intruders, 4 of which are spiral
galaxies, for instance. This indicates that compact radio sources,
although they are preferentially associated with red and massive ETGs,
can rarely be found in different hosts.

The {\em{WISE}} infrared colors further support the general passive
nature of the \FRo\ hosts. \FRo\ and \FR\ sources, see
Fig. \ref{wise}, show mid-IR colors typical of those of elliptical
galaxies \citep{wright10}. In particular, the \FRo\ overlaps with the
FR~I hosts at similarly low redshift. Nonetheless, a few galaxies of
the \FRo\ extend to redder colors than those from the \FR. They do not
follow the BL~Lacs sequence \citep{massaro12}, but reach the locus of
spiral and star-forming galaxies. Of the five sources with
$W2-W3> 2.5$, four are those noted above as late-type galaxies, with
blue optical colors, and/or they host black holes
$M_{\rm BH} \leq 10^{7.8} M_\odot$: they confirm the presence of a
minority of compact radio galaxies that do not conform with the
general properties of FR~0s.

\citet{ledlow96} discovered that FR~Is and FR~IIs populate
well-separated regions in a plane, based on the host optical
luminosity versus their radio power. Owing to their low radio
luminosities, FR~0s are well below the boundary between the two
classical FR classes, precisely, they are lower by a factor 10 to
200 (see Fig. \ref{ledlow}).

\begin{figure*}
\includegraphics[width=9.2cm]{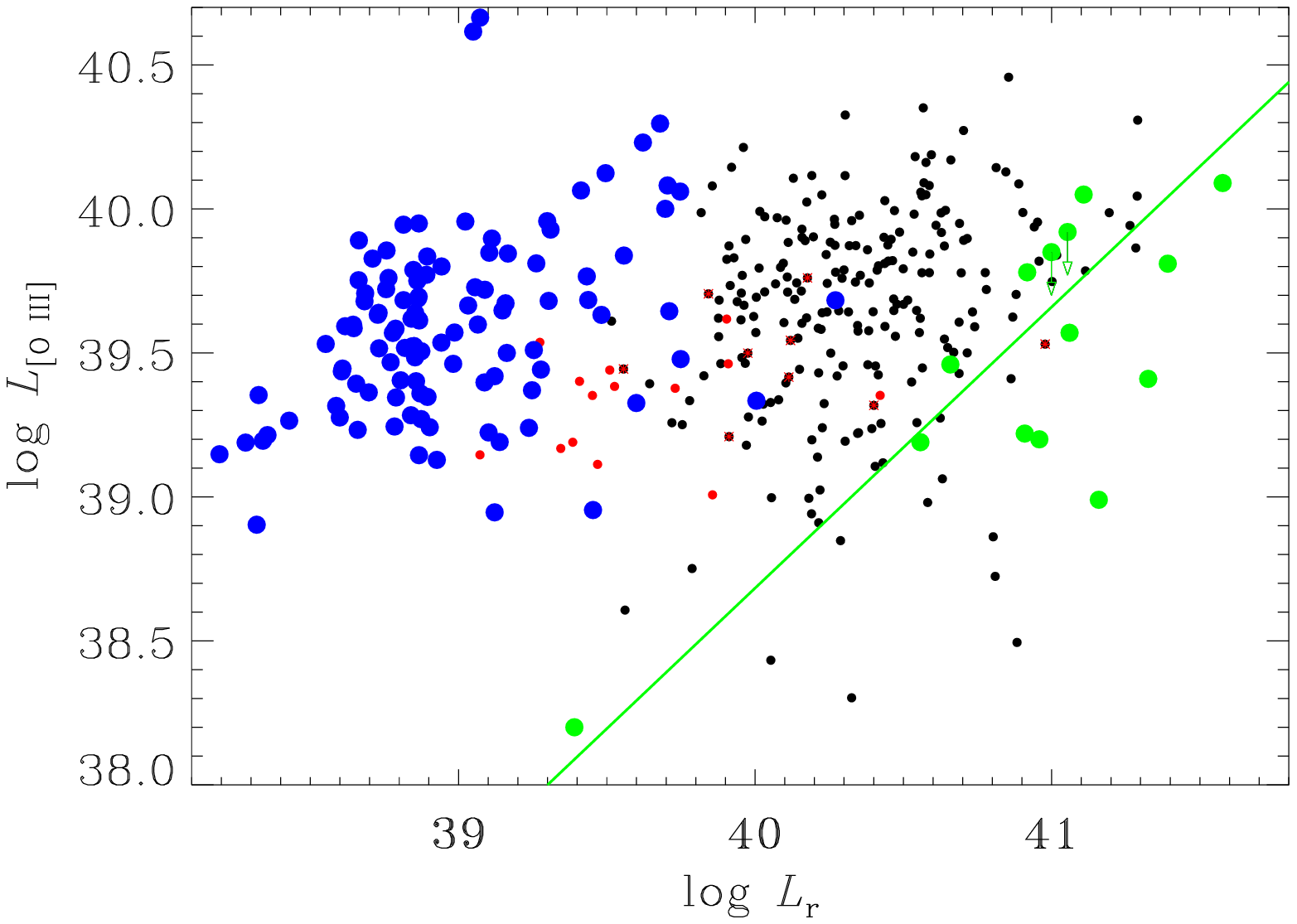}
\includegraphics[width=9.2cm]{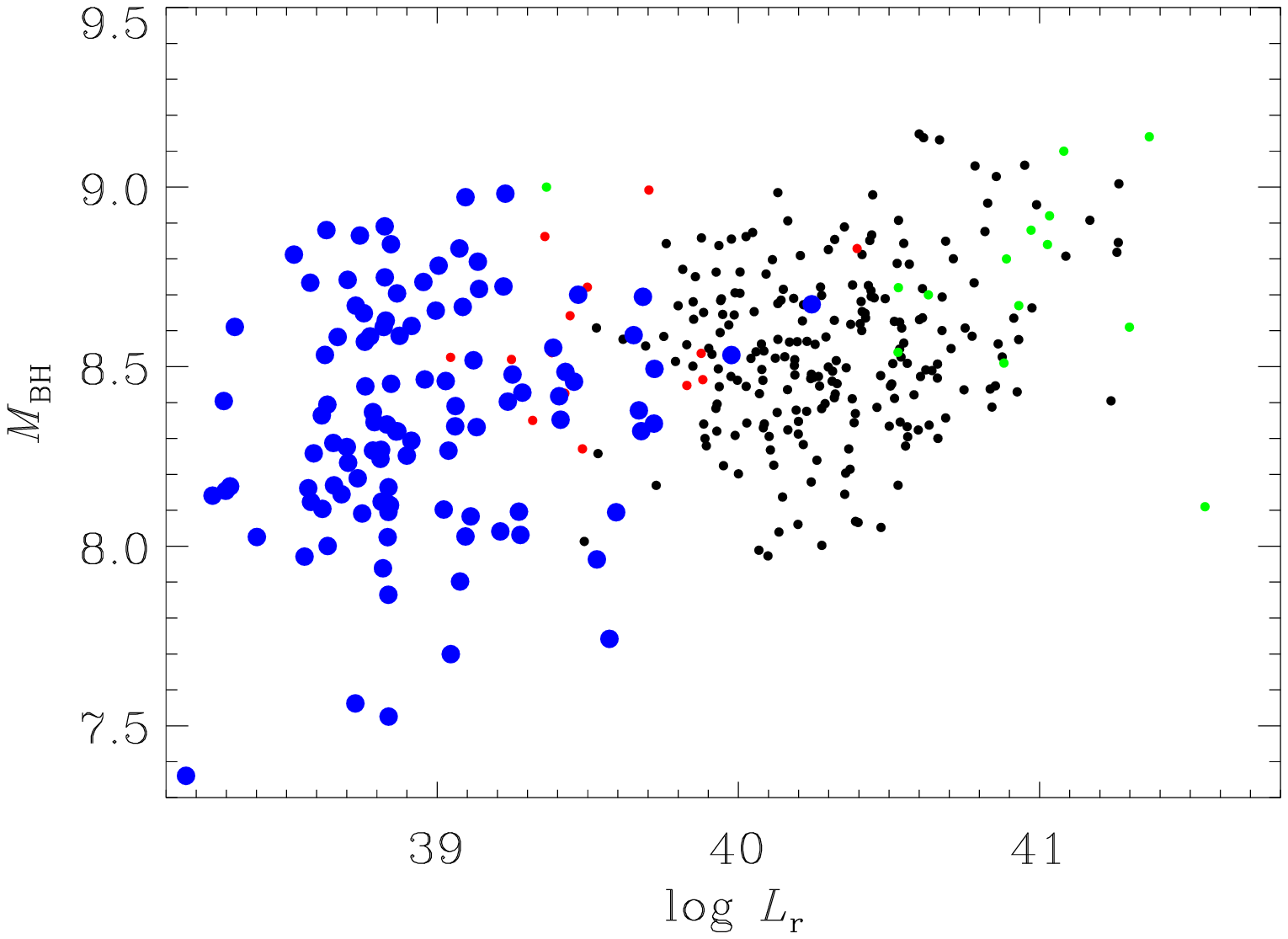}
\caption{Left panel: radio (NVSS) versus [O~III] line luminosity of the
  \FRo\ (blue), \FR\ (black), and \sFR\ (red) sources. The \FR\ sources with
  $z<0.05$ are represented as black dots with a red asterisk
superposed. The
  green line shows the correlation between these two quantities derived from
  the FR~Is of the 3C sample from \citep{buttiglione10}, individually marked
  with the green crosses. Right panel: radio luminosity versus BH
  mass (M$_{\odot}$).}
\label{mrur}
\end{figure*}

The line luminosity is a robust proxy of the radiative power of the AGN,
and at least within objects with similar multiwavelength properties,
of the accretion rate. The comparison of line emission and radio power
presented in Fig. \ref{mrur} (left panel) indicates that FR~0s share
the same range of $L_{\rm [O~III]}$ of FR~Is, but they have a much
lower radio luminosity, with a median 30 smaller than that for the
\FR.  Low-luminosity RGs form a continuous distribution, from the
FR~0s at the lowest ratios of radio/line luminosity, to the \sFR\ and
\FR\ sources at intermediate ratios, and finally to the extreme
3C-FRIs.  However, we note that the number of objects in this figure is
not proportional to their space density because they are selected in
different volumes. A proper comparison can be drawn only between FR~0s
and FR~Is with $z<0.05$, marked with the red dots for those with
$10<r<30$ kpc and with red asterisks when $r>30$ kpc. Figure~\ref{mrur}
(left panel) also points to a similarity between FR~0s and FR~Is in
terms of AGN bolometric power and accretion rate. An analogous
indication was found by \citet{baldi15} when comparing the genuine
core emission of FR~0s and FR~Is.

The right panel of Fig.~\ref{mrur} shows the relation between the radio
luminosity and the BH mass for the \FRo, the \FR, and the \sFR. The inclusion
of low-power RGs breaks down the relation between $L_{r}$ an $M_{BH}$ observed
at higher radio luminosities \citep{lacy01}. RGs with similar BH masses can
produce sources of different radio morphologies/sizes and spanning several orders of
magnitude in radio luminosity. Furthermore, a clearly empty region is present in
the lower right region of the diagram, indicating that a minimum BH mass is
required to launch a relativistic jet for a RLAGN.

\section{Discussion}

\subsection{Comparison with previous studies.}

Our study follows several previous investigations of the local
population of compact radio sources. In particular, \citet{sadler14}
explored the properties of nearby galaxies (z $\lesssim 0.1$) at 20
GHz, finding similar to our results, that compact radio sources are
the dominant class. However, compared to our sample, they found a more
heterogeneous population of hosts, including $\sim$ 25\% of HEGs
(while our HEGs fraction is only $\sim$ 4\%), and an indication of a
mixture of several types of objects, possibly also including young
GHz peaked-spectrum, compact steep-spectrum, and beamed
sources. The most likely origin of these differences is the higher
flux limit (40 mJy) and observing frequency (20 GHz), which leads to
the selection of more luminous objects, with compact sources reaching
radio luminosities as high as $10^{42}$ $\ergs$, which is a factor 100
higher than the most luminous FR~0s in our catalog. As we mentioned
above, this luminosity difference also characterizes the study of
compact sources performed by \citet{miraghaei17}.

Conversely, our FR~0s are in general more luminous than the compact
radio sources associated with nearby ETGs. For example, in the study
from \citet{wrobel91b} only a handful of objects reaches
$\sim 10^{38}$ $\ergs$, which is the luminosity limit of our
sample. The multiwavelength study of these 'miniature radiogalaxies'
(Core galaxies, \citealt{balmaverde06core}) shows various similarities
with FR~0s, including the host properties, the nuclear LEG spectrum,
and the high core dominance.

  From this perspective, our study fills the gap between the
  population of brighter radio galaxies and the weak radio sources in
  nearby ETGs: FR~0s are more similar to their less luminous
  counterparts.

\subsection{Are FR~0s young radio galaxies?}

The relative number density between FR~0s and FR~Is can be used to
explore the relation between these two classes of RGs. If we assume
that these two classes of low-luminosity RGs are linked by temporal
evolution, they are intrinsically identical sources differing
only by their age. This idea is supported by the strong similarity of
the host properties of the \FRo\ and \FR\ sources.

In this scenario, RGs are initially compact FR~0s and then they all
expand to form well-developed radio jets and appear as FR~I radio
sources.  This means that FR~0s are ``young" with respect to the time
needed to form FR~I RGs, whose sizes often reach 100 kpc, that is,
$\sim 10^8 \,v_{3}^{-1}$ years (where $v_{3}$ is the expansion speed
in 10$^3$ km s$^{-1}$ units). By assuming a constant expansion speed,
the relative space densities of the various classes are expected to be
proportional to the range of sizes covered by each class. In our case,
the space densities should scale with the linear size as $<$5 : 20 :
30 (kpc) for the FR~0s, the small FR~Is (with $10 < r < 30$ kpc), and
the larger FR~Is (with $30 < r < 60$ kpc). The observed numbers are
108 : 14 : 7.\footnote{The space density of the more extended FR~Is
  (the only class visible at larger distances) is confirmed by the
  number of FIRST/FRIs found with $z<0.15$, 158 sources within a
  volume 25 times larger.} The space density of FR~0s is then
$\gtrsim$50 times higher than predicted by this simple model.

These results could be explained, still in the evolution framework, if
the expansion speed increases with time, being higher for the sources
with a larger size. Nevertheless, this sharp acceleration contrasts
with the results from numerical simulations of low-luminosity radio
jets \citep{massaglia16} that show an approximately constant advance
speed out to a few kpc from the center and then a {\sl decrease} by a
factor $\sim$2.

This discrepancy is not significantly reduced by excluding the FR~0s
with host properties that do not conform with those of the FR~Is
discussed in Sect. 3 (i.e., objects having
$M_{\rm BH} < 10^8 M_\odot$, $C_r < 2.6$, or blue colors); when we
remove these likely interlopers, the FR~0s density drops by only
$\sim$15\%.

On the other hand, the sizes of edge-darkened FR~I radio sources that
we used above to predict the relative space densities are generally
underestimated because their extended plumes and tails can be detected
only to the sensitivity limit of the radio images. This effect would
decrease the genuine number of small FR~Is, further increasing the
FR~0/FR~I number density ratio. An opposite effect is due to
projection, since the observed size of extended radio sources is
reduced on average by a factor 2 with respect to its actual value,
assuming a random orientation.

\citet{capetti17a} discussed various sources of incompleteness of the
\FR\ sample. The most important limitation in our ability to detect
FR~Is is related to the presence of diffuse emission, which is
resolved out or does not reach the 3$\sigma$ limit in the FIRST
images. The \FR\ flux distribution indicates that indeed its
completeness limit is higher than the original 5 mJy threshold and can
be set at $\sim$30-50 mJy. When we restrict the comparison between
FR~0 and FR~I to the sources above 30 mJy, the FR~0/FR~I fraction
decreases, but by less than a factor $\sim$2.  Conversely, we expect
that the luminosity of a growing radio source increases with time as a
result of the contribution from the extended emission.  This causes an
underestimate of the intrinsic relative densities of FR~0 and extended
radio sources when using flux-limited samples.

\citet{capetti17b} showed that the hosts of the majority of
edge-brightened radio sources (FR~IIs) consist of massive ETGs,
similar to those of FR~I and FR~0. We then explored the effect of
dropping the requirement that FR~0s evolve only into FR~Is and considered
all extended sources regardless of their radio morphology.  In the
BH12 catalog there are 203 radio AGN with $z<0.05$, $M_{\rm BH} > 10^8
M_\odot$ that are spectroscopically identified as LEGs; 25 of them have
$10 < r < 30$ kpc, and 14 extend over $30 < r < 60$ kpc. This reduces
the difference between the observed and predicted number of compact radio
sources by only a factor $\sim$2.

To summarize, although the measurement of the relative number density
of the various classes of RGs is subject to various uncertainties, a
simple ``age-size'' scenario of the evolution of FR~0s into extended
RGs cannot be reconciled with the properties of such a class, meaning
that FR~0s cannot just be ``young'' RGs that will all eventually evolve
into extended radio sources.

\subsection{Are radio galaxies recurrent?}

The activity in RGs might be recurrent (e.g.,
\citealt{reynolds97,czerny09}) and they might experience various
active phases of different length. To account for both the large
fraction of compact radio sources and the presence of $\sim$100 kpc
scale sources, the duration of the active phase must cover a wide
range of values with short active periods favored toward the longer
ones. In this scenario, FR~0s are indeed young RGs, but they will
generally not grow to form large RGs. The limit to the size of FR~0s
corresponds to an age of $\lesssim 5\times 10^6 v_{3}^{-1}$ years.

Another constraint on recurrence comes from the bivariate radio/optical
luminosity functions. \citet{mauch07} estimated that the fraction of galaxies
associated with a radio source more luminous than $5 \times 10^{38} \ergs$ at
1.4 GHz (the lower luminosity seen in FR~0s) is $\sim$ 10\% for hosts with
$M_K \lesssim -25$, a fraction that decreases to $\sim$2\% for less luminous
galaxies, $-24 \gtrsim M_K \gtrsim -25$. When we adopt a typical color $R-K=2.7$
\citep{mannucci01}, these values encompass the range of optical magnitudes
of most FR~0s. These fractions can be interpreted within a recurrence scenario
as the ratio of the duration between the active and quiescent states, the
latter being between one and two orders of magnitude longer, that is, $\lesssim
10^{7}-10^{8}$ years.

In this context, it is important to explore the origin of the
recurrence and what sets its timescale; it can be envisaged that the
AGN feedback plays an important role in this respect (see, e.g.,
\citealt{pellegrini12}). A possible clue comes from the slightly
higher luminosity (and BH mass) of FR~Is with respect to FR~0s, and
the lack of compact sources in galaxies brighter than $M_r \sim -23$.
This suggests that the activity in more massive galaxies might last
for longer periods. This agrees with the results obtained by
\citet{shin12}, with cluster central galaxies spending a much longer
time in the active states than satellite galaxies.

\subsection{Are FR~0s and FR~Is intrinsically different?}

There might also be genuine differences between FR~0s and FR~Is either
in their central engines or in their environment, to account for their
distinct radio behavior.

The immediate environment of FR~0s and FR~Is is probed by their host
galaxies, which, as we discussed, are very similar: this argues
against the interpretation that the FR~0 hosts have a denser
interstellar medium that prevents their jets from propagating through
the galaxy. The large-scale environment could have a direct effect on
the radio source appearance because an extended confining medium
reduces the advance speed of the jet and also the adiabatic expansion
of the radio source, hence increases their detectability. However,
this is unlikely to be the case because the FR~0s size limit, $<$5
kpc, locates these sources well within the hot coronae associated with
their host. Furthermore, \citet{miraghaei17} concluded that FR~Is and
compact RGs live in a similar environment; given the differences
between our FR~0 definition and their compact RGs sample, a detailed
study of the FR~0 environment is needed before we can extend this
conclusion to our sample.

Based on high-resolution JVLA maps \citet{baldi15} suggested that the
differences between compact and extended sources might be due to a
lower jet bulk speed, $\Gamma_{\rm jet}$, in FR~0s than in FR Is. For
this reason they are more likely to be subject to instabilities and
entrainment \citep{bodo13} and they disrupt while they slowly burrow
their way into the external medium, which accounts for their small
sizes. This is supported by the absence of one-sided kpc scale
morphologies, a sign of relativistic jet boosting, in
FR~0s. Conversely, more than half of the FR~Is show large jet
asymmetries with a jet/counter-jet flux ratio higher than 2
\citep{parma87}, which implies that their jets are relativistic on a
kpc scale.

The possible lower $\Gamma_{\rm jet}$ in FR~0 does not appear to be
related to a (currently) directly observable
quantity. \citeauthor{baldi15} speculated that this could be due to a
positive connection between $\Gamma_{\rm jet}$ and the BH spin (as
suggested by \citealt{mckinney05}, \citealt{tchekhovskoy10},
\citealt{chai12}, \citealt{maraschi12}). In this context, when we
assume a galaxy evolution via BH merger and gas accretion (e.g.,
\citealt{volonteri13}), the result of such a process is a population
with a broad distribution of properties, that is, BH mass and spin. We
suggest that only under the most favorable circumstances, that is,
when these parameters are maximized, is the BH associated with an
RLAGN able to launch a highly relativistic jet and produce an extended
RG, classified as FR~I/II. This scenario predicts a large scatter of
physical BH aspects, which reproduces the different radio morphologies
and shapes a continuous population distribution of RGs from FR~0s to
FR~I/IIs. FR~0s would originate from less extreme values of the BH
parameters and represent the majority of the RLAGN population.

\section{Summary and conclusions}

We explored the properties of compact radio sources associated with
low-redshift ($z <0.05$) galaxies. We selected these objects by
cross-matching the SDSS, FIRST, and NVSS catalogs with a radio flux
$>$5 mJy, a radio size $\lesssim$ 5 kpc, and an LEG optical
spectrum. We named these sources FR~0s, in contrast to the other
extended FR classes, and formed the \FRo\ catalog. With these
criteria, we selected 108 objects with the following properties: 1)
massive red early-type hosts associated with massive BH
($\gtrsim$10$^{8}$ M$_{\odot}$), as typical of RLAGN
\citep{baldi10b,best12,chiaberge11}, and similar to those seen in \FR\
and 3C-FRIs; 2) a tail toward slightly lower $M_{\rm BH}$ values is
present in FR~0s; 3) there is a notable lack of blue host galaxies in
the \FRo, already observed in \FR, with respect to the general
population of ETGs \citep{schawinski09}. The available data do not
allow us to identify the origin of this effect, which might be due to
a different environment or to AGN feedback, for example.

From our study we are able to estimate the number density of the
different classes of RGs among the three FR-{\it CAT} samples. In the
same volume we identify 108 FR~0s and 21 FR~Is (14 with sizes of
between 10 and 30 kpc, and 7 between 30 and 60 kpc), and just one
FR~II, indicating that FR~0s are the dominant population of RGs in the
local Universe. This indicates that FR~0s cannot be RGs viewed
preferentially at a small angle with respect to their jet axis.

The high fraction of compact RGs rules out the possibility that FR~0s
are young RGs that will all evolve into extended radio sources. Two
alternatives to account for both the similarities and differences
between compact and extended RGs remain: 1) radio activity might be
recurrent, with short active periods favored with respect to the
longer ones. In this scenario, FR~0s are indeed young RGs, but they
will generally not grow to form large RGs; 2) there might be intrinsic
differences in their central engines, for example, compact radio
sources might be associated with jets of lower bulk speed than the
extended ones. Slower jets are more likely to be subject to
instabilities and entrainment and they disrupt before they exit to the
host interstellar medium, limiting the extension of the resulting
radio source. The possible ultimate origin of this low jet bulk speed
is a lower BH spin than that in FR~I/IIs.

The connection between galaxy and radio properties might depend on the
environment because the galaxy richness of the surroundings might lead
to different evolutionary paths. In particular the specific history of
mergers determines not only the evolution of the hosts but also of the
BH parameters, leading to the differentiation between FR~0 and FR~I
sources. The availability of the \FRo\ and \FR\ samples opens the
possibility for a detailed comparison of the environment of these two
classes of RGs.

Gathering the results of our studies of the three FR-{\it CAT} samples
of radio galaxies in the local Universe, we conclude that
low-luminosity RGs are a very homogeneous population of sources: all
hosted in red massive elliptical galaxies with an LEG spectrum, and
confined within a small range of emission line luminosities. Combined
with the information available on their nuclear properties
\citep{baldi15}, this suggests that a common jet-launching mechanism
operates in all these objects. However, they show large differences in
radio behavior from the point of view of their sizes, luminosities,
and morphologies. FR~0s and FR~I/IIs represent the two extremes in
terms of radio sizes and luminosities. Nonetheless, despite the
selection criteria (i.e., the separation between compact and extended
radio sources), there is no indication of dichotomous behavior: in
particular, the distribution in radio luminosity of FR~0s smoothly
connects with that of FR~I and FR~II.

Clearly, the classification of a given source in the FR classes
depends on sensitivity, resolution, and frequency of the available
radio data. For example, two of the FR~0s observed at high resolution
($\sim$ 0\farcs2) by \citet{baldi16} show an FR~I morphology. It can
be envisaged that future radio surveys at high resolution will resolve
most of the compact FR~0s. Nevertheless, we note that FR~0s are not
simply the scaled-down versions of FR~Is. The two classes differ not
only from the point of view of their sizes, but also by their radio
core dominance and by the ratio between line and radio luminosity,
which are both higher in FR~0s. This points to a genuine physical
difference, that is, a reduced ability in the production of extended
radio emission.

The increasing interest in low-luminosity AGN with the forthcoming
advent of the SKA array \citep{whittam17} makes the study of this
population of compact weak radio galaxies deserving of a deeper
investigation. Progress in the study of compact RGs can come from
further studies at high spatial resolution of FR~0 to explore their
jet bulk speed. This can be done in particular by measuring the
jet/counter-jet asymmetries in a sizable sample of FR~0s whose
distribution is directly linked to the jet speed. This will be
addressed in forthcoming papers that explore FR~0s at higher
resolutions with the JVLA and eMERLIN radio arrays. We also expect
that low-frequency radio observations from the LOFAR array will shed
new light on the possible recurrent activity in RGs in general.

\bibliographystyle{aa} 
\bibliography{./my.bib} 

\begin{acknowledgements}
  We thank the anonymous reviewer for his/her comments and suggestions
  to improve the interpretations of the results. The authors also thank
  D. Williams for reading the manuscript and providing useful
  comments.  RDB acknowledges the support of STFC under grant
  [ST/M001326/1].

This research made use of the NASA/ IPAC Infrared Science Archive and
Extragalactic Database (NED), which are operated by the Jet Propulsion
Laboratory, California Institute of Technology, under contract with the
National Aeronautics and Space Administration.

Funding for SDSS-III has been provided by the Alfred P. Sloan Foundation, the
Participating Institutions, the National Science Foundation, and the
U.S. Department of Energy Office of Science. The SDSS-III web site is
http://www.sdss3.org/.  SDSS-III is managed by the Astrophysical Research
Consortium for the Participating Institutions of the SDSS-III Collaboration,
including the University of Arizona, the Brazilian Participation Group,
Brookhaven National Laboratory, University of Cambridge, Carnegie Mellon
University, University of Florida, the French Participation Group, the German
Participation Group, Harvard University, the Instituto de Astrofisica de
Canarias, the Michigan State/Notre Dame/JINA Participation Group, Johns
Hopkins University, Lawrence Berkeley National Laboratory, Max Planck
Institute for Astrophysics, Max Planck Institute for Extraterrestrial Physics,
New Mexico State University, New York University, Ohio State University,
Pennsylvania State University, University of Portsmouth, Princeton University,
the Spanish Participation Group, University of Tokyo, University of Utah,
Vanderbilt University, University of Virginia, University of Washington, and
Yale University.

\end{acknowledgements}

\onecolumn
\begin{center}
\begin{longtable}{l l r r l l l l l c l c}

\caption[Properties of the FR~0s candidates sample.]{Properties of the FR~0s candidates sample.} 
\label{tab} \\

\hline \hline 
&  \,\,\,\,\,z & NVSS & [O~III] & \,\,\,\,\,m$_{\rm r}$ & \,\,Dn & \,\,\,$\sigma_*$ & \,\,\,C$_{\rm r}$ & \,\,\,\, $\nu L_r$ &  $L_{\rm[O~III]}$ & $M_{\rm BH}$ \\
\hline	
\endfirsthead

\multicolumn{3}{c}{{\tablename} \thetable{} -- Continued} \\[0.5ex]
\hline \hline 
& \,\,\,\,\,z & NVSS & [O~III] & \,\,\,\,\,m$_{\rm r}$ & \,\,Dn & \,\,\,$\sigma_*$ & \,\,\,C$_{\rm r}$ & \,\,\,\, $\nu L_r$ &  $L_{\rm[O~III]}$ & $M_{\rm BH}$ \\
\hline
\endhead

\hline
  \multicolumn{10}{c}{{Continued on Next Page}} \\
\endfoot

  \\[-1.8ex] 
\endlastfoot

SDSS~J010852.48-003919.4 & 0.045 &   10.9 &  115.7 & 15.095 &   2.00 & 223 &   3.43 &  38.89 &  39.77 &   8.3 \\
SDSS~J011204.61-001442.4 & 0.044 &   17.9 &   51.2 & 14.836 &   1.93 & 225 &   2.78 &  39.09 &  39.40 &   8.3 \\
SDSS~J011515.78+001248.4 & 0.045 &   42.6 &   84.3 & 14.554 &   1.93 & 241 &   3.10 &  39.48 &  39.63 &   8.5 \\
SDSS~J015127.10-083019.3 & 0.018 &   35.7 &  267.6 & 13.351 &   1.97 & 183 &   3.03 &  38.59 &  39.32 &   8.0 \\
SDSS~J020835.81-083754.8 & 0.034 &   28.4 &  186.8 & 13.694 &   1.93 & 242 &   2.97 &  39.06 &  39.73 &   8.5 \\
SDSS~J075354.98+130916.5 & 0.048 &    7.4 &   51.5 & 14.347 &   2.01 & 305 &   3.36 &  38.77 &  39.47 &   8.9 \\
SDSS~J080716.58+145703.3 & 0.029 &   28.4 &   63.3 & 13.712 &   1.97 & 215 &   3.38 &  38.93 &  39.13 &   8.3 \\
SDSS~J083158.49+562052.3 & 0.045 &    9.0 &   93.1 & 14.514 &   1.99 & 216 &   2.96 &  38.81 &  39.68 &   8.3 \\
SDSS~J083511.98+051829.2 & 0.046 &   10.1 &   60.5 & 14.495 &   1.93 & 241 &   3.24 &  38.87 &  39.51 &   8.5 \\
SDSS~J084102.73+595610.5 & 0.038 &    8.9 &  111.9 & 14.000 &   1.97 & 229 &   3.28 &  38.64 &  39.60 &   8.4 \\
SDSS~J084701.88+100106.6 & 0.048 &   23.7 &   48.4 & 14.508 &   1.90 & 244 &   3.22 &  39.28 &  39.44 &   8.5 \\
SDSS~J090652.79+412429.7 & 0.027 &   51.8 &  143.9 & 13.808 &   1.90 & 189 &   2.81 &  39.12 &  39.42 &   8.0 \\
SDSS~J090734.91+325722.9 & 0.049 &   46.9 &   35.0 & 14.972 &   1.49 & 160 &   2.41 &  39.60 &  39.33 &   7.7 \\
SDSS~J090937.44+192808.2 & 0.028 &   69.1 &  342.2 & 13.877 &   1.82 & 234 &   3.29 &  39.26 &  39.81 &   8.4 \\
SDSS~J091039.92+184147.6 & 0.028 &   50.0 &   78.9 & 13.298 &   1.89 & 195 &   2.86 &  39.14 &  39.19 &   8.1 \\
SDSS~J091601.78+173523.3 & 0.029 &   24.5 &  266.1 & 13.091 &   1.91 & 225 &   2.35 &  38.86 &  39.75 &   8.3 \\
SDSS~J091754.25+133145.5 & 0.050 &   22.9 &   76.3 & 15.603 &   1.93 & 189 &   3.11 &  39.30 &  39.68 &   8.0 \\
SDSS~J093003.56+341325.3 & 0.042 &   33.1 &  192.2 & 14.358 &   1.97 & 237 &   3.14 &  39.31 &  39.93 &   8.4 \\
SDSS~J093346.08+100909.0 & 0.011 &   56.6 &  567.0 & 12.095 &   2.05 & 204 &   3.02 &  38.34 &  39.20 &   8.2 \\
SDSS~J093938.62+385358.6 & 0.046 &    6.1 &   74.3 & 14.864 &   1.96 & 197 &   3.10 &  38.65 &  39.59 &   8.1 \\
SDSS~J094319.15+361452.1 & 0.022 &   75.1 &  583.4 & 13.132 &   1.81 & 175 &   3.15 &  39.10 &  39.85 &   7.9 \\
SDSS~J100549.83+003800.0 & 0.021 &   24.1 &  321.5 & 13.604 &   2.01 & 296 &   3.36 &  38.55 &  39.53 &   8.8 \\
SDSS~J101329.65+075415.6 & 0.046 &    7.8 &  137.0 & 14.333 &   2.04 & 272 &   3.21 &  38.76 &  39.86 &   8.7 \\
SDSS~J101806.67+000559.7 & 0.048 &   14.3 &  781.8 & 14.965 &   1.57 & 156 &   3.04 &  39.07 &  40.66 &   7.7 \\
SDSS~J102403.28+420629.8 & 0.044 &    6.0 &   39.9 & 14.652 &   1.86 & 204 &   3.17 &  38.60 &  39.28 &   8.2 \\
SDSS~J102511.50+171519.9 & 0.045 &   10.2 &   61.1 & 13.890 &   2.00 & 309 &   3.26 &  38.85 &  39.48 &   8.9 \\
SDSS~J102544.22+102230.4 & 0.046 &   76.7 &   57.6 & 14.405 &   1.93 & 246 &   3.02 &  39.75 &  39.48 &   8.5 \\
SDSS~J103719.33+433515.3 & 0.025 &  132.2 &  326.4 & 13.226 &   1.92 & 227 &   3.35 &  39.44 &  39.68 &   8.4 \\
SDSS~J103952.47+205049.3 & 0.046 &    6.9 &  126.6 & 14.380 &   1.91 & 202 &   3.14 &  38.71 &  39.83 &   8.1 \\
SDSS~J104028.37+091057.1 & 0.019 &   68.5 &  376.4 & 12.528 &   1.93 & 220 &   3.12 &  38.94 &  39.54 &   8.3 \\
SDSS~J104403.68+435412.0 & 0.025 &   32.4 &  126.1 & 13.427 &   1.86 & 213 &   3.27 &  38.84 &  39.28 &   8.2 \\
SDSS~J104811.90+045954.8 & 0.034 &   49.1 &  313.2 & 13.597 &   1.82 & 196 &   2.99 &  39.30 &  39.96 &   8.1 \\
SDSS~J104852.92+480314.8 & 0.041 &   19.2 &  990.9 & 14.183 &   1.77 & 197 &   2.87 &  39.05 &  40.62 &   8.1 \\
SDSS~J105731.16+405646.1 & 0.025 &   44.8 &  189.5 & 12.916 &   2.00 & 283 &   3.08 &  38.98 &  39.46 &   8.7 \\
SDSS~J111113.18+284147.0 & 0.029 &   41.1 &  196.5 & 13.480 &   1.92 & 216 &   2.89 &  39.07 &  39.60 &   8.3 \\
SDSS~J111622.70+291508.2 & 0.045 &   71.5 &   86.0 & 13.773 &   2.00 & 276 &   2.91 &  39.71 &  39.64 &   8.7 \\
SDSS~J111700.10+323550.9 & 0.035 &   17.6 &  298.7 & 13.905 &   1.86 & 204 &   3.28 &  38.87 &  39.95 &   8.2 \\
SDSS~J112029.23+040742.1 & 0.050 &    7.5 &  142.0 & 14.276 &   1.97 & 230 &   2.41 &  38.81 &  39.95 &   8.4 \\
SDSS~J112039.95+504938.2 & 0.028 &   24.5 &  136.5 & 14.251 &   1.96 & 259 &   3.22 &  38.80 &  39.40 &   8.6 \\
SDSS~J112256.47+340641.3 & 0.043 &   16.6 &  199.3 & 13.477 &   2.01 & 270 &   3.23 &  39.02 &  39.96 &   8.7 \\
SDSS~J112625.19+520503.5 & 0.048 &    9.0 &   85.2 & 15.336 &   1.93 & 172 &   3.39 &  38.87 &  39.70 &   7.9 \\
SDSS~J112727.52+400409.4 & 0.035 &   13.8 &  178.4 & 14.749 &   1.94 & 144 &   3.48 &  38.76 &  39.72 &   7.6 \\
SDSS~J113446.55+485721.9 & 0.032 &   13.2 &  101.1 & 13.924 &   1.94 & 252 &   3.27 &  38.65 &  39.39 &   8.5 \\
SDSS~J113449.29+490439.4 & 0.033 &   33.0 &   61.2 & 13.183 &   2.00 & 299 &   3.33 &  39.10 &  39.22 &   8.8 \\
SDSS~J113637.14+510008.5 & 0.050 &    9.0 &   35.7 & 14.818 &   1.93 & 223 &   3.39 &  38.90 &  39.35 &   8.3 \\
SDSS~J114230.94-021505.3 & 0.047 &    8.8 &   73.8 & 14.385 &   1.92 & 216 &   3.04 &  38.84 &  39.62 &   8.3 \\
SDSS~J114232.84+262919.9 & 0.030 &   42.0 &   39.3 & 13.028 &   2.02 & 324 &   3.20 &  39.12 &  38.95 &   9.0 \\
SDSS~J114804.60+372638.0 & 0.042 &   29.1 &   54.0 & 13.818 &   1.89 & 281 &   3.18 &  39.25 &  39.37 &   8.7 \\
SDSS~J115531.39+545200.4 & 0.050 &   31.2 &   94.1 & 14.869 &   1.96 & 236 &   3.16 &  39.43 &  39.77 &   8.4 \\
SDSS~J120551.46+203119.0 & 0.024 &   89.9 &  126.5 & 13.776 &   1.94 & 190 &   3.27 &  39.24 &  39.24 &   8.0 \\
SDSS~J120607.81+400902.6 & 0.037 &    9.5 &  163.4 & 13.636 &   2.00 & 233 &   3.17 &  38.66 &  39.75 &   8.4 \\
SDSS~J121329.27+504429.4 & 0.031 &   96.5 &  574.2 & 12.825 &   1.87 & 277 &   3.04 &  39.50 &  40.12 &   8.7 \\
SDSS~J121951.65+282521.3 & 0.027 &    8.7 &   46.8 & 14.247 &   2.06 & 234 &   3.18 &  38.32 &  38.90 &   8.4 \\
SDSS~J122421.31+600641.2 & 0.044 &    6.1 &   57.7 & 14.043 &   2.04 & 283 &   3.46 &  38.61 &  39.44 &   8.7 \\
SDSS~J123011.85+470022.7 & 0.039 &   93.8 &  264.0 & 13.570 &   1.91 & 231 &   2.94 &  39.70 &  40.00 &   8.4 \\
SDSS~J124318.73+033300.6 & 0.048 &   63.5 &  211.5 & 14.310 &   1.94 & 223 &   3.13 &  39.71 &  40.08 &   8.3 \\
SDSS~J124633.75+115347.8 & 0.047 &   61.2 &  354.5 & 14.021 &   1.90 & 260 &   3.09 &  39.68 &  40.30 &   8.6 \\
SDSS~J125027.42+001345.6 & 0.047 &   54.5 &  310.1 & 15.380 &   1.73 & 196 &   3.35 &  39.62 &  40.23 &   8.1 \\
SDSS~J125409.12-011527.1 & 0.047 &    7.7 &   66.6 & 14.743 &   1.92 & 196 &   3.30 &  38.78 &  39.57 &   8.1 \\
SDSS~J130404.99+075428.4 & 0.046 &   10.5 &  128.2 & 13.618 &   1.97 & 278 &   3.18 &  38.89 &  39.84 &   8.7 \\
SDSS~J130837.91+434415.1 & 0.036 &   58.4 &  366.3 & 13.438 &   1.89 & 255 &   2.89 &  39.41 &  40.06 &   8.6 \\
SDSS~J133042.51+323249.0 & 0.034 &   17.9 &  218.5 & 14.256 &   1.85 & 179 &   3.30 &  38.85 &  39.79 &   7.9 \\
SDSS~J133455.94+134431.7 & 0.023 &   39.4 &  335.7 & 12.854 &   2.02 & 285 &   3.18 &  38.85 &  39.64 &   8.7 \\
SDSS~J133621.18+031951.0 & 0.023 &   30.4 &  258.8 & 13.280 &   1.96 & 212 &   3.22 &  38.73 &  39.52 &   8.2 \\
SDSS~J133737.49+155820.0 & 0.026 &   26.9 &  235.8 & 12.968 &   2.02 & 257 &   2.90 &  38.79 &  39.58 &   8.6 \\
SDSS~J134159.72+294653.5 & 0.045 &   10.4 &   81.8 & 14.463 &   1.85 & 196 &   3.02 &  38.87 &  39.62 &   8.1 \\
SDSS~J135036.01+334217.3 & 0.014 &  101.3 &  677.6 & 12.524 &   2.00 & 199 &   3.16 &  38.84 &  39.52 &   8.1 \\
SDSS~J135226.71+140528.5 & 0.023 &   25.5 &  133.5 & 12.971 &   2.07 & 307 &   3.40 &  38.66 &  39.23 &   8.9 \\
SDSS~J140528.32+304602.0 & 0.025 &    7.4 &   93.2 & 14.159 &   1.97 & 129 &   3.03 &  38.19 &  39.15 &   7.4 \\
SDSS~J141451.35+030751.2 & 0.025 &   26.7 &  370.8 & 13.025 &   1.88 & 207 &   3.13 &  38.76 &  39.76 &   8.2 \\
SDSS~J141517.98-022641.0 & 0.047 &   18.9 &  126.2 & 14.192 &   2.03 & 280 &   3.22 &  39.17 &  39.85 &   8.7 \\
SDSS~J142724.23+372817.0 & 0.032 &   20.8 &   72.5 & 13.721 &   2.05 & 301 &   3.43 &  38.87 &  39.27 &   8.8 \\
SDSS~J143156.59+164615.4 & 0.048 &    8.7 &   42.7 & 13.956 &   2.04 & 266 &   3.27 &  38.86 &  39.40 &   8.6 \\
SDSS~J143312.96+525747.3 & 0.047 &   15.6 &   93.3 & 15.076 &   1.57 & 232 &   3.35 &  39.09 &  39.72 &   8.4 \\
SDSS~J143424.79+024756.2 & 0.028 &    7.3 &   82.4 & 14.106 &   1.99 & 201 &   3.23 &  38.28 &  39.19 &   8.1 \\
SDSS~J143620.38+051951.5 & 0.029 &   18.7 &  212.1 & 13.337 &   1.97 & 284 &   3.35 &  38.73 &  39.64 &   8.7 \\
SDSS~J144745.52+132032.2 & 0.044 &    6.7 &  158.1 & 15.126 &   1.93 & 186 &   3.10 &  38.66 &  39.89 &   8.0 \\
SDSS~J145216.49+121711.5 & 0.031 &    8.0 &   76.7 & 14.222 &   1.94 & 188 &   3.23 &  38.43 &  39.26 &   8.0 \\
SDSS~J145243.25+165413.4 & 0.046 &   17.5 &  149.2 & 14.001 &   1.94 & 272 &   3.17 &  39.11 &  39.90 &   8.7 \\
SDSS~J145616.20+203120.6 & 0.045 &   25.8 &   65.1 & 13.922 &   1.98 & 326 &   3.35 &  39.25 &  39.51 &   9.0 \\
SDSS~J150152.30+174228.2 & 0.047 &   18.6 &   85.0 & 14.411 &   1.92 & 224 &   3.00 &  39.16 &  39.67 &   8.3 \\
SDSS~J150425.68+074929.7 & 0.049 &    7.8 &   54.5 & 14.979 &   2.04 & 226 &   3.13 &  38.82 &  39.52 &   8.3 \\
SDSS~J150601.89+084723.2 & 0.046 &    8.3 &   41.8 & 14.225 &   1.93 & 240 &   3.30 &  38.79 &  39.35 &   8.4 \\
SDSS~J150636.57+092618.3 & 0.028 &   27.8 &   73.9 & 14.329 &   1.68 & 141 &   2.94 &  38.87 &  39.15 &   7.5 \\
SDSS~J150808.25+265457.6 & 0.033 &   20.3 &   87.4 & 15.185 &   2.00 & 198 &   3.54 &  38.87 &  39.36 &   8.1 \\
SDSS~J152010.94+254319.3 & 0.034 &   18.3 &  120.9 & 13.756 &   2.05 & 263 &   3.41 &  38.85 &  39.52 &   8.6 \\
SDSS~J152151.85+074231.7 & 0.044 &   11.7 &   35.8 & 13.855 &   2.02 & 260 &   2.85 &  38.90 &  39.24 &   8.6 \\
SDSS~J153016.15+270551.0 & 0.033 &   13.3 &  195.8 & 14.309 &   1.85 & 205 &   3.08 &  38.69 &  39.71 &   8.2 \\
SDSS~J154147.28+453321.7 & 0.037 &    8.9 &  117.6 & 14.094 &   1.91 & 215 &   3.44 &  38.62 &  39.59 &   8.3 \\
SDSS~J154426.93+470024.2 & 0.038 &   17.6 &  177.5 & 13.642 &   1.98 & 264 &   3.11 &  38.94 &  39.80 &   8.6 \\
SDSS~J154451.23+433050.6 & 0.037 &   11.5 &  129.3 & 13.646 &   1.92 & 217 &   3.27 &  38.73 &  39.63 &   8.3 \\
SDSS~J155951.61+255626.3 & 0.045 &  144.7 &   43.3 & 14.516 &   1.89 & 252 &   3.18 &  40.00 &  39.33 &   8.5 \\
SDSS~J155953.99+444232.4 & 0.042 &   59.5 &  158.9 & 14.474 &   1.89 & 182 &   3.26 &  39.56 &  39.84 &   8.0 \\
SDSS~J160426.51+174431.1 & 0.041 &   96.0 &  276.4 & 15.416 &   1.89 & 226 &   2.32 &  39.75 &  40.06 &   8.3 \\
SDSS~J160523.84+143851.6 & 0.041 &    8.6 &   55.8 & 13.705 &   2.01 & 259 &   3.04 &  38.70 &  39.36 &   8.6 \\
SDSS~J160616.02+181459.8 & 0.037 &  396.0 &  143.1 & 13.507 &   1.78 & 273 &   3.24 &  40.27 &  39.68 &   8.7 \\
SDSS~J160641.83+084436.8 & 0.047 &    9.3 &   87.1 & 14.452 &   1.93 & 188 &   2.51 &  38.86 &  39.69 &   8.0 \\
SDSS~J161238.84+293836.9 & 0.032 &   27.4 &  147.3 & 14.036 &   1.93 & 242 &   3.44 &  38.99 &  39.57 &   8.5 \\
SDSS~J161256.85+095201.5 & 0.017 &   21.7 &  323.8 & 12.856 &   2.07 & 203 &   3.39 &  38.33 &  39.35 &   8.2 \\
SDSS~J162146.06+254914.4 & 0.048 &    9.1 &   70.9 & 14.168 &   1.95 & 198 &   2.94 &  38.87 &  39.61 &   8.1 \\
SDSS~J162549.96+402919.4 & 0.029 &   97.5 &   43.3 & 13.121 &   1.97 & 245 &   2.91 &  39.45 &  38.95 &   8.5 \\
SDSS~J162846.13+252940.9 & 0.040 &   25.2 &  111.4 & 14.281 &   1.96 & 250 &   3.39 &  39.15 &  39.65 &   8.5 \\
SDSS~J162944.98+404841.6 & 0.029 &    7.7 &   77.9 & 16.542 &   1.98 & 263 &   2.14 &  38.36 &  39.21 &   8.6 \\
SDSS~J164925.86+360321.3 & 0.032 &   11.9 &  114.4 & 14.121 &   1.99 & 199 &   3.23 &  38.61 &  39.45 &   8.1 \\
SDSS~J165830.05+252324.9 & 0.033 &   13.1 &  181.9 & 14.326 &   1.93 & 219 &   3.43 &  38.68 &  39.68 &   8.3 \\
SDSS~J170358.49+241039.5 & 0.031 &   32.7 &  196.3 & 13.369 &   2.02 & 290 &   3.33 &  39.03 &  39.66 &   8.8 \\
SDSS~J171522.97+572440.2 & 0.027 &   57.2 &  174.0 & 12.560 &   1.97 & 292 &   2.81 &  39.16 &  39.50 &   8.8 \\
SDSS~J172215.41+304239.8 & 0.046 &    8.1 &   32.7 & 13.689 &   2.05 & 269 &   3.07 &  38.78 &  39.24 &   8.6 \\

\hline
\hline
\end{longtable}
\end{center}
Column description: (1) source name; (2) redshift; (3) NVSS 1.4 GHz
flux density [mJy]; (4) [O~III] flux [in 10$^{-17}$ erg cm$^{-2}$
s$^{-1}$ units]; (5) SDSS DR7 r band AB magnitude; (6) concentration
index $C_r$; (7) Dn(4000) index; (8) stellar velocity dispersion
[\kms]; (9) logarithm of the NVSS radio luminosity [erg s$^{-1}$];
(10) logarithm of the [O~III] line luminosity [erg s$^{-1}$]; (11)
logarithm of the black hole mass [in solar units].  \twocolumn

\end{document}